\documentclass[11pt,a4paper]{article}
\usepackage[hyperref]{acl2020}
\usepackage{times}
\usepackage{latexsym}
\usepackage{graphicx}

\usepackage{microtype}

\aclfinalcopy % Uncomment this line for the final submission
\usepackage{hyperref}

\makeatletter
\newcommand{\printfnsymbol}[1]{%
  \textsuperscript{\@fnsymbol{#1}}%
}
\makeatother

\title{Textual Analysis of Communications in COVID-19 Infected Community on Social Media}

\author{Yuhan Liu\thanks{~~ Equal contribution.}, Yuhan Gao\printfnsymbol{1}, Zhifan Nan\printfnsymbol{1}, Long Chen\printfnsymbol{1}\\
  New York University\\
  \texttt{\{yl7576,yg2417,zn2041,lc3424\}@nyu.edu} \\
 }

\date{}

\begin{document}
\maketitle

\begin{abstract}
During the COVID-19 pandemic, people started to discuss about pandemic-related topics on social media. On subreddit \textit{r/COVID19positive}, a number of topics are discussed or being shared, including experience of those who got a positive test result, stories of those who presumably got infected, and questions asked regarding the pandemic and the disease. In this study, we try to understand, from a linguistic perspective, the nature of discussions on the subreddit. We found differences in linguistic characteristics (e.g. psychological, emotional and reasoning) across three different categories of topics. We also classified posts into the different categories using SOTA pre-trained language models. Such classification model can be used for pandemic-related research on social media.
\end{abstract}

\section{Introduction}

Starting in late 2019, the COVID-19 pandemic has rapidly impacted over 200 countries, areas, and territories. As of December 7th, 2020, according to the World Health Organization (WHO), 66,243,918 COVID-19 cases were confirmed worldwide, with 1,528,984 confirmed deaths\footnote{https://covid19.who.int}. This disease has tremendous impacts on people’s daily lives worldwide.

With the pandemic spreading in the United States, people who tested positive started sharing information about their physical condition, emotion and story with the virus. In addition, those who have not gotten infected are curious about the symptoms and nature of the virus, as well as procedures of testing services across the country. A community of those who want to share their own story and who want to know more about the virus emerged on Reddit, a platform for any user (older than 13 years) to discuss, connect, and share their experiences and opinions online. Under subreddit \textit{r/COVID19positive}, people are sharing and discussing the virus, while seeking and giving emotional supports. An online community like this can have mixed emotions and splendid textual contents.

In this study, we investigate the linguistic features of contents in the subreddit. First, we classify the threads into different categories, including a) reporting of positive COVID-19 case, b) reporting of a presumed COVID-19 case, and c) question regarding COVID-19. Second, we aim to investigate linguistic characteristics of posts and subsequent comments in different contexts. Specifically, we found differences in contents when people are posting for different purposes (i.e. self-report their discussion to be in one of the three aforementioned categories).

\section{Related Work}

A large number of studies were performed with LIWC, an API\footnote{https://liwc.wpengine.com/} for linguistic analysis of documents. \citet{tumasjan2010predicting} used LIWC to capture the political sentiment and predict elections with Twitter. The API was also used by \citet{zhang2020contributes} to provide insights into the sentiment of the descriptions of crowdfunding campaigns.

Previous studies have also attempted to make textual classification on social media data. \citet{mouthami2013sentiment} implemented a classification model that approximately classifies the sentiment using Bag of words in Support Vector Machine (SVM) algorithm. \citet{huang2014cyber} applied SMOTE (Synthetic Minority Oversampling TEchnique) method to defecting online cyber-bullying behavior. In addition, a number of other studies performed textual classifications for various purposes using social media data~\citep{chen2020eyes, chatzakou2015harvesting, lukasik2016hawkes}.

\section{Dataset}

Data from subreddit \textit{r/COVID19positive} between March 14, 2020\footnote{The date when the subreddit was created.} and October 14, 2020 is collected using Pushshift API \footnote{https://www.pushshift.io.}. In total, 17,285 submissions (contents that starts a Reddit thread) were collected. As a medium-sized subreddit with 91.1K members\footnote{As of October 14, 2020.}, contents in this community should contain limited fake posts or misinformation, therefore leading to a relatively clean dataset.

Submission on Reddit starts a discussion with a title and an optional textual body. The title and the body are naturally good source for textual analysis. In addition, most submissions have \textit{flair}, a hashtag-like, user-reported label that describes the category of discussion under which the submission is about. The \textit{flairs} serve as a perfect label for potential supervised classification tasks. 

Data cleanup and preprocessing are preformed with the collected dataset. First, all posts without flairs are deleted. This left 15,410 posts in the dataset. Next, title and body are concatenated into one field, \textit{titletext}, so that the new field can be used as textual input for models. Then, we removed emojis\footnote{Since emojis are not supported by some pre-trained models, we removed them for consistency.}, extra separators and repeated punctuations from the text. Lastly, since we have 10 different flairs but limited dataset size, we merged related flairs into one category, resulting in three categories: a) question, b) tested\_positive, and c) presumed\_positive, as shown in Table~\ref{table:1}.

\begin{table}[htbp]
\vspace{0.3cm}
\caption{Categorization of flairs.}
\vspace{-0.1cm}
\small
\begin{center}
\begin{tabular}{r| c}
\hline
\textbf{Category} & \textbf{Original Flairs}\\
\hline
Question&Question\\
($N=8,687$)&Question - to those who tested positive\\
&Question - for medical research\\
&Medical question\\
\hline
Tested&Tested Positive\\
Positive&Tested Positive - Me\\
($N=5,241$)&Tested Positive - Family\\
&Tested Positive - Friends\\
\hline
Presumed&Presumed Positive - from doctor\\
Positive&Presumed Positive - from test\\
($N=1,482$)&\\

\hline
\end{tabular}
\label{table:1}
\end{center}
\vspace{-0.2cm}
\end{table}

\section{Methodology}

Two natural language processing tasks are conducted in order to investigate sentiments and linguistic characteristics of the Reddit posts, and to make classifications of different submissions, as more details explained below.

\subsection{Analysis of Linguistic Characteristics}

In this exploratory task, we aim to find discrepancies of texts with topics in different categories. Therefore, Linguistic Inquiry and Word Count (LIWC) is applied to extract the sentiment of submissions and comment of our corpus. LIWC2015 is a dictionary-based linguistic analysis tool that can count the percentage of words that reflect different emotions, thinking styles, social concerns, and capture people's psychological states\footnote{https://liwc.wpengine.com/how-it-works/}. We concatenated post in the three categories respectively to form three big documents, as performed by~\citet{yu2008exploring}, then use LIWC to get scores for each of the categories. Significance tests are performed to find relevant fields. We also manually selected field of interests even if no significant differences are found. In the end, we selected 3 summary linguistic features (\textit{Analytics}, \textit{Clout} and \textit{Tone}), 5 psychological features (\textit{pos\_emo}, \textit{neg\_emo}, \textit{sad}, \textit{anxiety} and \textit{anger}) and 3 time-oriented features (\textit{focuspast}, \textit{focuspresent} and \textit{focusfuture}).

\subsection{Classification}

With the categorized flairs, we build three-label classification models. A number of models were used, with details explained below.

\subsubsection{Stacking Ensemble Model}

We build a stacking ensemble model, with Random Forest, SVM, Naïve Bayes, XGBoost, Logistic Regression and K-Nearest Neighbor as meta-models. To transform our dataset into model-compatible format, we apply term frequency–inverse document frequency (TF-IDF) vectorization process on the dataset. Default hyperparameters are used.

\subsubsection{Bi-LSTM}

Next, we build a Bidirectional Long-Short Term Memory (Bi-LSTM) model. The dataset was converted into a 50-dimension word vector using spaCy\footnote{https://spacy.io} encoding. Hyperparameters were tuned with grid search method, with best ones as: \textbf{ADAM} optimizer, \textbf{lr=}1e-3, \textbf{eps=}1e-8 and \textbf{dropout=}0.2. Best performance was achieved at epoch=5.

\subsubsection{BERT}

\textbf{B}idirectional \textbf{E}ncoder \textbf{R}epresentations from \textbf{T}ransformers~\citep{devlin2018bert} is used as our first pre-trained language model. We used the BERT-base with cased input model, as more complex models performed poorly on our limited-sized dataset. We fine-tuned the model using a dense layer and an output layer of 3 neurons. Hyperparameters are also tuned using grid search, with best ones as: \textbf{ADAM} optimizer, \textbf{lr=}1e-5, \textbf{eps=}1e-8 and \textbf{hidden\_size=}50. Best performance was achieved at epoch=3

\subsubsection{XLNet}

XLNet~\citep{yang2019xlnet} is used as our second pre-trained language model. To keep comparability, we chose the XLNet-base with cased input model. We alsofine-tuned the model using a dense layer and an output layer of 3 neurons. Hyperparameters are tuned using grid search, with best ones as: \textbf{ADAM} optimizer, \textbf{lr=}3e-5, \textbf{eps=}1e-8 and \textbf{hidden\_size=}50. Best performance was achieved at epoch=4.

\subsubsection{Classification dataset}

Dataset was converted into model-compatible formats using various tokenization/vectorization methods. We then made a train-validation-test split with a 70:15:15 ratio. As the dataset is imbalanced among the three classes, we upsampled the minority classes in the training set using SMOTE~\citep{chawla2002smote}.

% First, we would like to build a baseline model with RNN with Word2Vec embeddings. Next, Bidirectional Long Short Term Memory (Bi-LSTM) model with Word2Vec is planned as a more advanced RNN model. In addition, we will attempt SOTA pre-trained language models with fine tuning, including BERT and XL-Net, aiming for better results. However, such models may struggle with limited corpus size of this study, as Ezen-Can points out in study~\cite{ezen2020comparison}.

\section{Results}

\subsection{Linguistic Characteristics}

\begin{figure}[htbp]
    \centering
    \includegraphics[width=\linewidth]{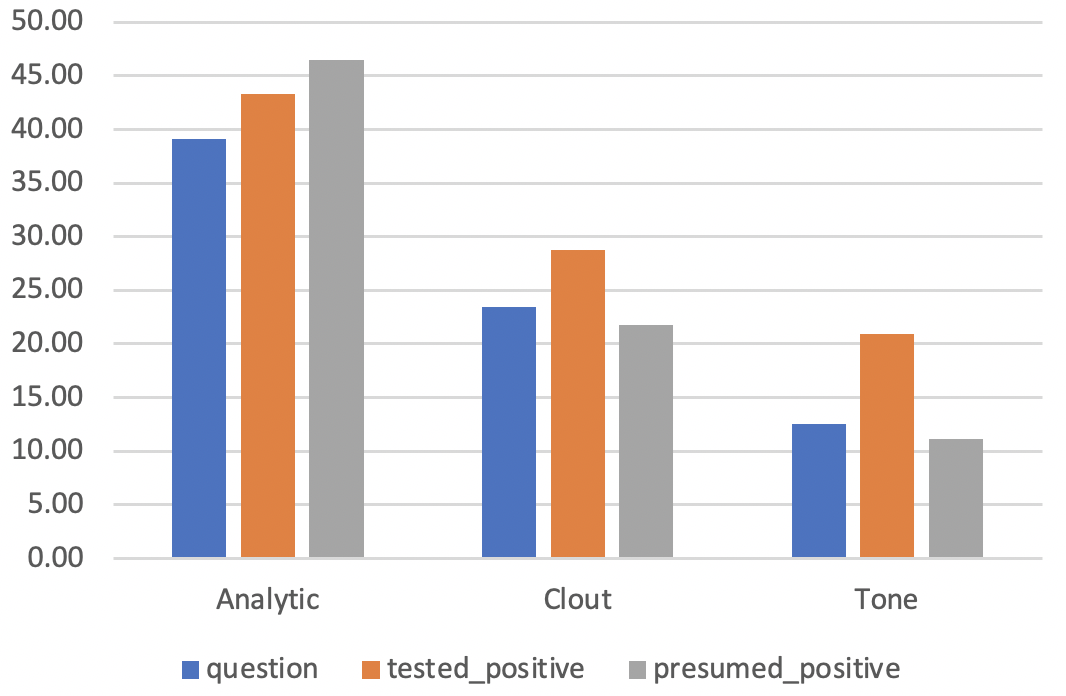}
      \vspace{-0.5cm}
    \caption{LIWC scores for Analytics, Clout and Tone. \textit{Presumed\_positive} posts have significantly higher Analytics score, while \textit{tested\_positive} posts have significantly higher Clout and Tone scores.}
    \label{fig:1}
    \vspace{-0.2cm}
\end{figure}

First, we found some differences in 3 summary linguistic features among the three classes, as shown in Fig.~\ref{fig:1}. \textit{Presumed\_positive} posts have higher Analytic (i.e. analytical thinking) score, inferring more logical and formal thinking presented in discussion. Indeed, most posts in presumed\_positive posts are posted by those who are very likely to be positive but still uncertain. Building upon uncertainty, they tend to start a logical analysis/reasoning on the symptoms and their recent activities which might get them infected. \textit{Question} posts have relatively lower analytic score, which can also be explained by the question-raising nature of such posts. In terms of Clout, which stands for the level of confidence, \textit{tested\_positive} posts have higher score on this feature, inferring that they are ``more certain'' about their positive diagnosis and their ``confidence'' of getting better. The same category of posts also has higher emotional tone scores, inferring that they are ``more emotional'' with getting infected.

\begin{figure}[htbp]
    \centering
    \includegraphics[width=\linewidth]{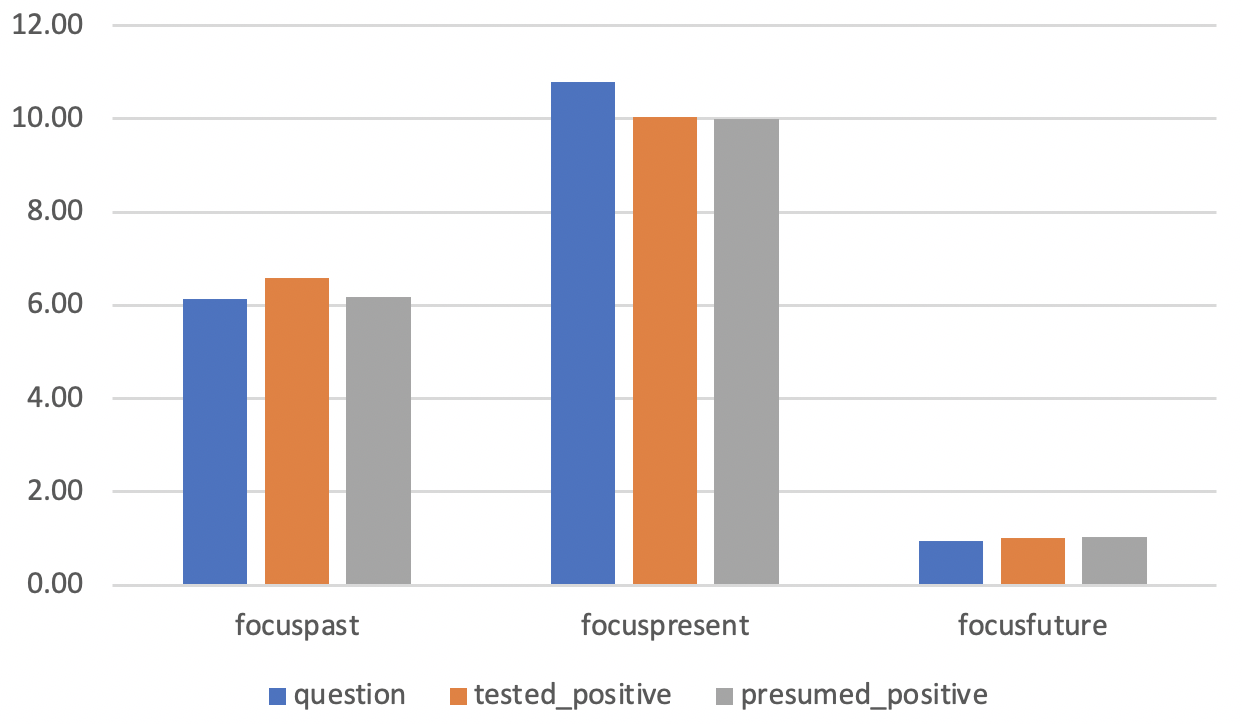}
      \vspace{-0.5cm}
    \caption{LIWC scores for time-oriented features. \textit{Tested\_positive} posts have significantly higher focus on the past, while \textit{question} posts have significantly higher focus on present. No significant differences found for focuses on future, and the levels of focus on future are low for all three classes.}
    \label{fig:2}
    \vspace{-0.2cm}
\end{figure}

Next, we investigate time-oriented features of the three categories of posts, as shown in Fig.~\ref{fig:2}. \textit{Tested\_positive} posts have higher focus on the past, while \textit{question} posts have more focus on present. By looking at sample posts, we found that \textit{tested\_positive} posts tend to reflect more on their potential cause of getting infected (e.g. hang out in a pub, attend a gathering without mask) and their symptoms before visiting a doctor, while \textit{question} posts tend to report their current feelings and symptoms, seeking for answers. In addition, all categories of posts does not have significant difference of focus on the future and that the magnitude of future-focus is low for the three categories comparing to focuses on past or present.

\begin{figure}[htbp]
    \centering
    \includegraphics[width=\linewidth]{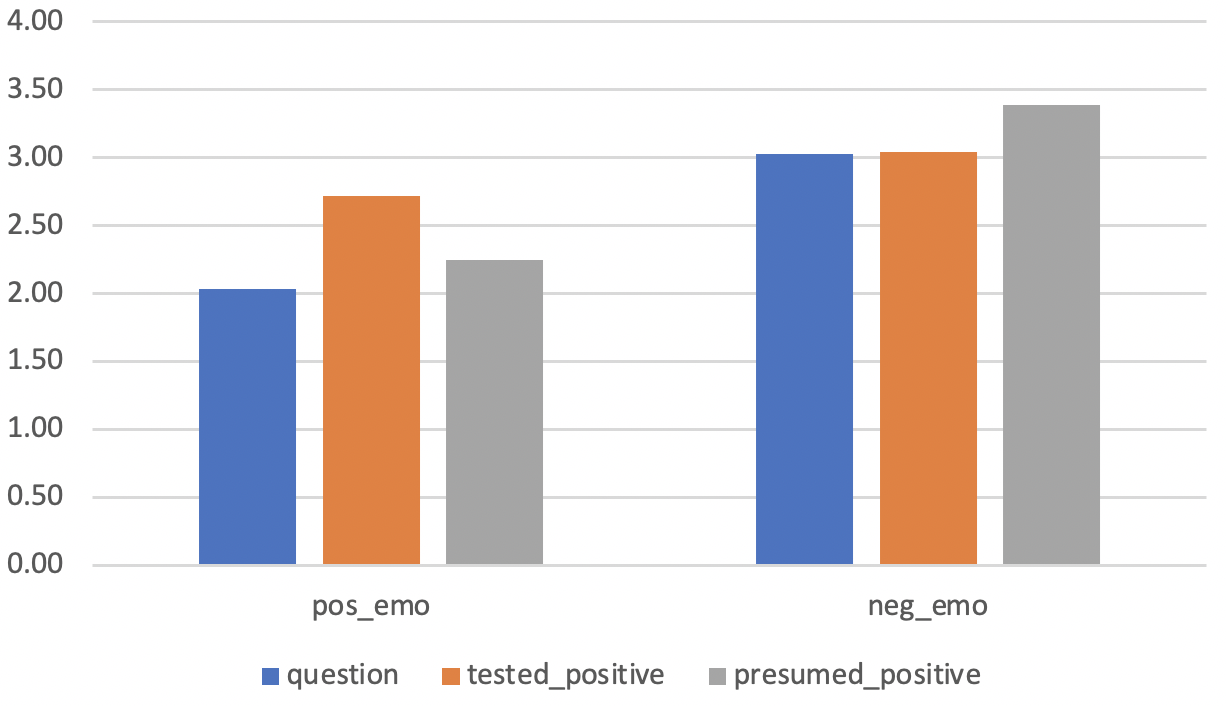}
      \vspace{-0.5cm}
    \caption{LIWC scores for psychological features. \textit{Tested\_positive} posts have significantly higher positive emotions, while \textit{presumed\_positive} posts have significantly higher negative emotions.}
    \label{fig:3}
    \vspace{-0.2cm}
\end{figure}

As for the emotional features, \textit{tested\_positive} posts have, surprisingly, higher level of positive emotions, inferring that those who got infected are more ``optimistic'' about getting better, while \textit{presumed\_positive} posts have higher level of negative emotions, which makes sense as they are the ones who are most uncertain and that are really feeling bad. As we look into different negative emotions, \textit{presumed\_positive} posts have significantly higher sadness level, which can be interpreted as depressed during uncertainty of a very likely diagnosis, while \textit{question} posts have significantly higher anxiety level. The high anxiety level from \textit{question} posts reflect their uncertainty and concerns about the pandemic in general.

\begin{figure}[htbp]
    \centering
    \includegraphics[width=\linewidth]{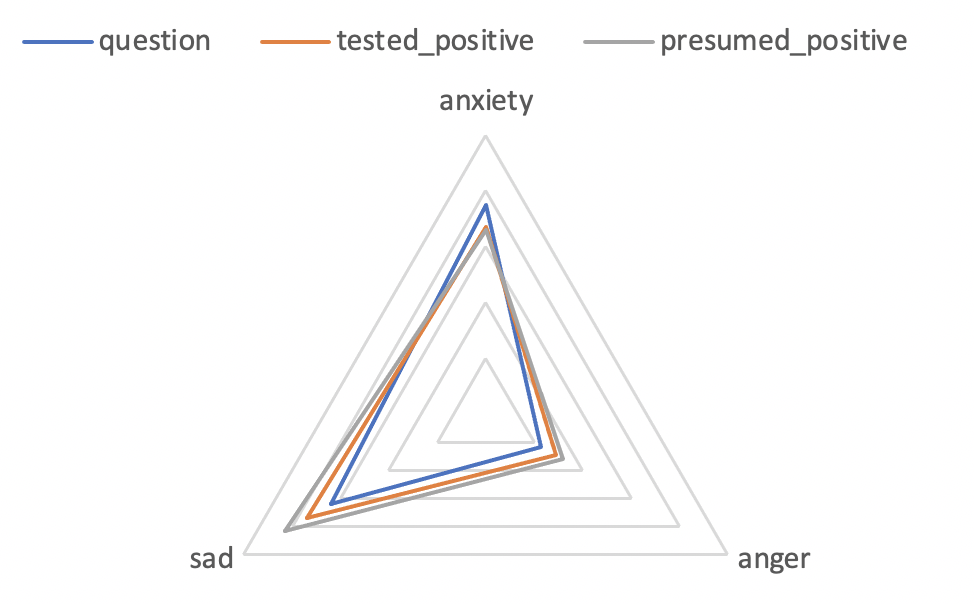}
      \vspace{-0.5cm}
    \caption{LIWC scores for specific negative emotions. \textit{Question} posts have significantly higher anxiety level, while \textit{presumed\_positive} posts have significantly higher sadness level. All three classes have low level of anger in comparison to the other two negative emotions.}
    \label{fig:4}
    \vspace{-0.2cm}
\end{figure}

\begin{table}[htbp]
\vspace{0.3cm}
\caption{Accuracy and F-1 scores for classification models.}
\vspace{-0.1cm}
\small
\begin{center}
\begin{tabular}{r| c c}
\hline
\textbf{Model} & \textbf{Accuracy} & F1\\
\hline
\textbf{Ensemble}&0.685 & 0.676\\
\textbf{Bi-LSTM}&0.625 & 0.604\\
\textbf{BERT}&\textbf{0.726} & \textbf{0.722}\\
\textbf{XLNet}&0.711 & 0.702\\

\hline
\end{tabular}
\label{table:2}
\end{center}
\vspace{-0.2cm}
\end{table}

\subsection{Classification}

The performance of classification models are shown in Table.~\ref{table:2}. The best model is BERT, with a testing F-1 score of 0.722. The model converges rather quickly, reaching best accuracy at epoch=3. XLNet achieved good performance as well. Stacking ensemble model also yields good result, with F-1 score of 0.676. However, Bi-LSTM model performed poorly, with F-1 score of only 0.604. We tried other encoding methods, including more embedding dimensions, without improvement.

We observed that the performance difference between SOTA pre-trained models and traditional models is not huge. We suspect that the relatively small size of dataset restricted better performance in BERT or XLNet, as these models are more complicated and thus require larger dataset to train on. Such finding is in congruent to the research by~\citet{ezen2020comparison}.

\section{Conclusion and Future Work}

In this study, we performed a linguistic analysis of posts in an online COVID-19 discussion community on Reddit. Posts in the three categories showed some differences with psychological, emotional and other characteristics. We also built classification models to differentiate posts among the categories, found that the SOTA pre-trained models yields best performance.

Future work can be done to incorporate more features into classification models, such as metadata from reddits (e.g. up/down votes, number of comments), and linguistic scores from LIWC model for each individual posts. Also, the classification model developed in our study can be used on social media to identify infected people for other studies, such as psychological evaluation of the infected group. In addition, as we only analyzed Reddit submissions, comments are the other textual source on Reddit with greater amount but without labels. In fact, more than 1 million comments were collected in this subreddit in comparison to only 17,285 submissions. Such data source can be for unsupervised machine learning for pattern recognition, such as Latent Dirichlet Allocation topic modelling.

\bibliographystyle{acl_natbib}
\bibliography{project}

\begin{thebibliography}{13}
\expandafter\ifx\csname natexlab\endcsname\relax\def\natexlab#1{#1}\fi

\bibitem[{Chatzakou and Vakali(2015)}]{chatzakou2015harvesting}
Despoina Chatzakou and Athena Vakali. 2015.
\newblock Harvesting opinions and emotions from social media textual resources.
\newblock \emph{IEEE Internet Computing}, 19(4):46--50.

\bibitem[{Chawla et~al.(2002)Chawla, Bowyer, Hall, and
  Kegelmeyer}]{chawla2002smote}
Nitesh~V Chawla, Kevin~W Bowyer, Lawrence~O Hall, and W~Philip Kegelmeyer.
  2002.
\newblock Smote: synthetic minority over-sampling technique.
\newblock \emph{Journal of artificial intelligence research}, 16:321--357.

\bibitem[{Chen et~al.(2020)Chen, Lyu, Yang, Wang, and Luo}]{chen2020eyes}
Long Chen, Hanjia Lyu, Tongyu Yang, Yu~Wang, and Jiebo Luo. 2020.
\newblock In the eyes of the beholder: Sentiment and topic analyses on social
  media use of neutral and controversial terms for covid-19.
\newblock \emph{arXiv preprint arXiv:2004.10225}.

\bibitem[{Devlin et~al.(2018)Devlin, Chang, Lee, and
  Toutanova}]{devlin2018bert}
Jacob Devlin, Ming-Wei Chang, Kenton Lee, and Kristina Toutanova. 2018.
\newblock Bert: Pre-training of deep bidirectional transformers for language
  understanding.
\newblock \emph{arXiv preprint arXiv:1810.04805}.

\bibitem[{Ezen-Can(2020)}]{ezen2020comparison}
Aysu Ezen-Can. 2020.
\newblock A comparison of lstm and bert for small corpus.
\newblock \emph{arXiv preprint arXiv:2009.05451}.

\bibitem[{Huang et~al.(2014)Huang, Singh, and Atrey}]{huang2014cyber}
Qianjia Huang, Vivek~Kumar Singh, and Pradeep~Kumar Atrey. 2014.
\newblock Cyber bullying detection using social and textual analysis.
\newblock In \emph{Proceedings of the 3rd International Workshop on
  Socially-Aware Multimedia}, pages 3--6.

\bibitem[{Lukasik et~al.(2016)Lukasik, Srijith, Vu, Bontcheva, Zubiaga, and
  Cohn}]{lukasik2016hawkes}
Michal Lukasik, PK~Srijith, Duy Vu, Kalina Bontcheva, Arkaitz Zubiaga, and
  Trevor Cohn. 2016.
\newblock Hawkes processes for continuous time sequence classification: an
  application to rumour stance classification in twitter.
\newblock In \emph{Proceedings of the 54th Annual Meeting of the Association
  for Computational Linguistics (Volume 2: Short Papers)}, pages 393--398.

\bibitem[{Mouthami et~al.(2013)Mouthami, Devi, and
  Bhaskaran}]{mouthami2013sentiment}
K~Mouthami, K~Nirmala Devi, and V~Murali Bhaskaran. 2013.
\newblock Sentiment analysis and classification based on textual reviews.
\newblock In \emph{2013 international conference on Information communication
  and embedded systems (ICICES)}, pages 271--276. IEEE.

\bibitem[{Tumasjan et~al.(2010)Tumasjan, Sprenger, Sandner, and
  Welpe}]{tumasjan2010predicting}
Andranik Tumasjan, Timm~O Sprenger, Philipp~G Sandner, and Isabell~M Welpe.
  2010.
\newblock Predicting elections with twitter: What 140 characters reveal about
  political sentiment.
\newblock In \emph{Fourth international AAAI conference on weblogs and social
  media}.

\bibitem[{Yang et~al.(2019)Yang, Dai, Yang, Carbonell, Salakhutdinov, and
  Le}]{yang2019xlnet}
Zhilin Yang, Zihang Dai, Yiming Yang, Jaime Carbonell, Russ~R Salakhutdinov,
  and Quoc~V Le. 2019.
\newblock Xlnet: Generalized autoregressive pretraining for language
  understanding.
\newblock In \emph{Advances in neural information processing systems}, pages
  5753--5763.

\bibitem[{Yu et~al.(2008)Yu, Kaufmann, and Diermeier}]{yu2008exploring}
Bei Yu, Stefan Kaufmann, and Daniel Diermeier. 2008.
\newblock Exploring the characteristics of opinion expressions for political
  opinion classification.
\newblock In \emph{Proceedings of the 2008 international conference on Digital
  government research}, pages 82--91. Digital Government Society of North
  America.

\bibitem[{Zhang et~al.(2016)Zhang, Li, Wang, and Zhou}]{zhang2016user}
Dong Zhang, Shoushan Li, Hongling Wang, and Guodong Zhou. 2016.
\newblock User classification with multiple textual perspectives.
\newblock In \emph{Proceedings of COLING 2016, the 26th International
  Conference on Computational Linguistics: Technical Papers}, pages 2112--2121.

\bibitem[{Zhang et~al.(2020)Zhang, Lyu, and Luo}]{zhang2020contributes}
Xupin Zhang, Hanjia Lyu, and Jiebo Luo. 2020.
\newblock What contributes to a crowdfunding campaign's success? evidence and
  analyses from gofundme data.
\newblock \emph{arXiv preprint arXiv:2001.05446}.

\end{thebibliography}
\nocite{*}

\end{document}